\documentclass[12pt,preprint,epsf]{aastex} 

\shorttitle{Six Open Clusters }
\shortauthors{Krisciunas \& Monteiro} 

\begin{document}
\received{06 November2017}

\title{Why is the Main Sequence of NGC 2482 So Fat?}   
\author{
Kevin Krisciunas,\altaffilmark{1}
Nicholas B. Suntzeff,\altaffilmark{1}
Daniel Q. Nagasawa,\altaffilmark{1}
Marshall C. Johnson,\altaffilmark{2}
William Cochran,\altaffilmark{3} and
Michael Endl\altaffilmark{3}
}
\altaffiltext{1}{George P. and Cynthia Woods Mitchell Institute for Fundamental 
Physics \& Astronomy, Texas A. \& M. University, Department of Physics \& Astronomy,
  4242 TAMU, College Station, TX 77843-4242; {krisciunas@physics.tamu.edu},
{nsuntzeff@tamu.edu}, {dqnagasawa@physics.tamu.edu} }

\altaffiltext{2}{Ohio State University, Department of Astronomy,
140 W. 18th Avenue, 4055 McPherson Laboratory, Columbus, OH 43210-1106;
{johnson.7240@osu.edu} }

\altaffiltext{3}{University of Texas, Austin, 2515 Speedway Stop C1402,
McDonald Observatory, Austin, TX 78712-1206; {wdc@astro.as.utexas.edu},
{mike@astro.as.utexas.edu} }

\begin{abstract} 


We present the results of high resolution spectra of seven stars in the 
field of NGC 2482, an open star cluster of age 447 Myr.  We confirm the 
previously published values of the radial velocity and metallicity of one 
giant star.  This gives us confidence that another giant star is a {\em bona 
fide} cluster member, and that three stars significantly above the main 
sequence in a color-magnitude diagram are {\em not} members, on the basis of 
discordant radial velocities.  Another star $\approx$1.7 mag above the main 
sequence may or may not be a member.  Its [Fe/H] value is $\sim$0.1 dex more 
positive than two giant stars studied, and its radial velocity is 3-4 km 
s$^{-1}$ less than that of the two giant stars, which is a significant 
difference if the velocity dispersion of the cluster is less than $\pm$1 km 
s$^{-1}$. To a large extent the width of the main sequence seems to be due 
to the presence of foreground and background stars in the same general direction, 
stars that masquerade as main sequence stars in the cluster.

\end{abstract}


\citet{Kri_etal15} give $uBVgri$ PSF photometry of 114 stars in the 
southern open cluster NGC 2482.  Table 2 of our paper gives the celestial 
coordinates of these stars and the star ID's we also use in the present 
paper.  From isochrone fits to the PSF photometry we find an age of the 
cluster of log $t$ = 8.65 $\pm$ 0.09, or $t$ = 447$^{+103} _{-84}$ Myr.
Until now only one star in this cluster has been 
studied spectroscopically \citep{Red_etal13}. This star, which we will 
refer to as star 9 of \citet{MV_75}, has a radial velocity of +39.00 
$\pm$ 0.2 km s$^{-1}$ and a metallicity [Fe/H] = log ($Z/Z_{\odot}$) = 
$-$0.07 $\pm$ 0.04, where $Z$ is the abundance of elements heavier than 
helium.  Presumably, this is the metallicity of the cluster.  Analysis
based solely on our photometry gives [Fe/H] = $-$0.10 (+0.13, $-$0.18).

Fig. \ref{cmd} shows a color-magnitude diagram of NGC 2482 based on a 
combination of short and long exposures, so as not to saturate the bright 
stars or to have too low a signal-to-noise ratio for the faint stars.  To 
Fig. \ref{cmd} we have added star 9 of \citet{MV_75}, using their single 
channel photoelectric photometry.  This star was outside our 8.7 by 8.7 
arcmin field of view. In Fig. \ref{cmd} stars 108 and 206 are at the 
cluster turnoff point. Star 23 is either a blue straggler or is situated 
beyond the cluster.  Star 166 and Moffat and Vogt's star 9, presumably 
{\em asymptotic} giants in the cluster, fall on an isochrone that fits the 
photometry of the upper main sequence stars and those at the turnoff point 
\citep{Kri_etal15}.

On the right hand side of Fig. \ref{cmd} the points plotted as magenta triangles 
clearly correspond to stars not in the cluster.  One might wonder: Why is 
the main sequence so fat?  What about the stars considerably above the 
main sequence? Unresolved binaries in the cluster might be up to 
2.5 log$_{10}$(2) $\approx$ 0.753 mag 
above the main sequence, but not more.  Could it be that this cluster's 
stars formed over a range of time, possibly with different abundances?  
That would be noteworthy, as we assume that the stars in an open cluster 
are coeval and have the same abundances.  Two stars in the Pleiades 
considerably above the main sequence are worthy of the reader's attention 
\citep{Opp_etal97}, as they have characteristics of stars much younger
than the cluster.   It is suggested that a cloud that was a member
of the ``Pleiades Supercluster'' recently formed stars, which are
now situated between us and the cluster.  These two stars could be members
of the group.  We know that some {\em globular} clusters show 
evidence for multiple main sequences, owing to different helium 
abundances \citep{Pio_etal07, Mil_etal12}, but the longer time scale for 
globular cluster evolution makes this easier.

To investigate the cluster membership of some stars in NGC 2482 we used 
the McDonald Observatory 2.7-m Harlan J. Smith Telescope and the 
cross-dispersed Tull coud\'{e} spectrograph, which provides a resolving 
power of 60,000 at 777 nm. Our data were taken on 15 and 16 March 2016 
UT.  As radial velocity standards we used HD 182572 and HD 42807.  
Given the declination of the cluster ($-$24\arcdeg \hspace{-1 mm}.3 ) 
and the latitude of McDonald Observatory, the cluster was high enough 
to observe about 5 hours per night.

The spectra were processed with an {\sc iraf}\footnote[4]{{\sc iraf} is 
distributed by the National Optical Astronomy Observatory, which is 
operated by the Association of Universities for Research in Astronomy, 
Inc., under cooperative agreement with the National Science Foundation 
(NSF).} script that performed standard bias subtraction, inter-order 
scattered light subtractions, flat fielding and optimal extraction from 
58 echelle orders.  We determined stellar parameters using the {\em 
Kea} software \citep{Endl_Cochran}, which matches selected portions of 
the observed spectrum against a library of model templates. These 
results are given in Table \ref{results}. 
Star 206, at the turnoff point of the cluster, is too hot and rapidly 
rotating to show sufficiently deep absorption lines of metals to give 
accurate values of the radial velocity and metallicity. As a sanity 
check, our new values of [Fe/H] and the radial velocity of star 9 of 
\citet{MV_75} match, within the errors, those previously published 
\citep{Red_etal13}.

We can assert that stars 200, 232, and 256 are not members of the 
cluster, as their radial velocities are 20 to 35 km s$^{-1}$ different 
than Moffat and Vogt's star 9 and our star 166, 
and the velocity dispersion of an open 
cluster is typically less than $\pm$1 km s$^{-1}$ \citep{Gel_etal10, 
Tof_etal14}. Our star 244 is a possible member of the cluster, as its 
radial velocity matches those of Moffat and Vogt star 9 and our star 166 
within 3-4 km s$^{-1}$.  Given that our star 166 and Moffat and Vogt's 
star 9 have almost identical $BV$ photometry, abundances, radial 
velocities, and temperatures, and given that they fall on an isochrone 
that matches the photometry of the cluster, we are very confident that 
they are both members of the cluster. Stars 206 and 108 (at the turnoff 
point) fall on the same isochrone.

We would like to have obtained spectra of other stars in NGC 2482 that 
do not neatly lie on the main sequence, but because of 
computer problems on the second of our half-nights, we were unable to 
observe more than a total of seven stars at a signal-to-noise ratio 
sufficiently high to make the analysis worthwhile. We note that $V$ 
$\approx$ 13.5 to 14 is about the limit with the 2.7-m telescope and 
the Tull spectrograph in 1 hr of total exposure time, in order to do 
abundance analysis and have confidence in cluster membership.

We note that the mean proper motion of NGC 2482 is only 3.3 
milliarcseconds per year in both right ascension and declination 
\citep[see Table 1 of][]{Kri_etal15}. The Gaia project should give us 
added information on the motions of stars in this cluster.

Still, we have confirmed the radial velocity and metallicity of star 9 of
\citet{MV_75}, and we have confirmed that that star and 
our star 166 are nearly identical regarding measurable parameters. Three 
stars observed by us to be above the main sequence in a CMD and in the 
direction of the cluster are clearly not cluster members, based on radial 
velocities.  Another star above the main sequence (star 244) may or may not be a 
cluster member.  Its metallicity [Fe/H] = +0.06 $\pm$ 0.04, which is
$\sim$0.1 dex more positive than the two giant stars studied.

\acknowledgments

We thank the McDonald Observatory Time Allocation Committee for the opportunity
to obtain our spectra.  We also thank Rebecca Oppenheimer for useful discussions.

\clearpage

\begin{deluxetable}{lcccccc}
\tablecolumns{7}
\tablewidth{0pc}
\tablecaption{Results from High Resolution Spectra\tablenotemark{a}\label{results}}
\tablehead{ \colhead{Star} & \colhead{RV (km s$^{-1}$)} & \colhead{$v$ sin $i$ (km s$^{-1}$)} &
\colhead{T$_{eff} (K)$} & 
\colhead{log $g$} & \colhead{[Fe/H]} & \colhead{Member?}  }
\startdata

MV star 9 & +38.7 $\pm$ 0.3 &  4.6 $\pm$ 0.2 & 4940 $\pm$ 98  & 2.38 $\pm$ 0.12 & $-$0.04 $\pm$ 0.03 & Y \\
166       & +37.4 $\pm$ 0.2 &  5.1 $\pm$ 0.2 & 4800 $\pm$ 45  & 2.25 $\pm$ 0.14 & $-$0.06 $\pm$ 0.03 & Y \\
206       & \ldots            & $>$ 60           & 9162 $\pm$ 143 & 4.89 $\pm$ 0.07 & $-$0.45 $\pm$ 0.20 & Y \\
244       & +34.6 $\pm$ 0.2 &  7.9 $\pm$ 0.2 & 6320 $\pm$ 86  & 4.38 $\pm$ 0.07 &   +0.06 $\pm$ 0.04 & Maybe \\
\\
200       & +57.4 $\pm$ 0.5 & 14.2 $\pm$ 0.3 & 6440 $\pm$ 125 & 3.62 $\pm$ 0.24 & $-$0.03 $\pm$ 0.05 & N \\
232       & +63.3 $\pm$ 0.7 &  8.8 $\pm$ 0.3 & 6860 $\pm$ 40  & 3.69 $\pm$ 0.64 & $-$0.40 $\pm$ 0.05 & N \\
256       & +72.2 $\pm$ 0.4 & 10.3 $\pm$ 0.2 & 6320 $\pm$ 169 & 4.31 $\pm$ 0.16 & $-$0.24 $\pm$ 0.04 & N \\
\enddata
\tablenotetext{a}{For each star we give the heliocentric radial velocity, 
the projected rotational speed, the effective 
temperature, log $g$ (where $g$ = GM/r$^2$, and log $g$ in cgs units is
$\approx$4.4 for the Sun), the metallicity, and whether the star is a member of the cluster.}
\end{deluxetable}

\figcaption[ngc2482_cmb_new.eps]{Color-magnitude diagram of NGC 2482.
Except for star number 9 of \citet{MV_75}, the photometry is
from Table 2 of \citet{Kri_etal15}; some stars are labelled according to
the numbering scheme of this paper. Blue circles correspond to presumed cluster
members observed with short (3 second) exposures on 6 January 2012 UT with the
Las Campanas Observatory 1-m Swope telescope.  Squares correspond to
presumed cluster members imaged with 150 second ($B$-band) and 120 second
($V$-band) exposures on 21 December 2012 UT.  Magenta triangles represent red stars
clearly not in the cluster.  The solid black line is the 447 Myr
isochrone from \citet{Kri_etal15}, which was fit to stars brighter than V = 15.5.
The red dashed line shows a portion of that
isochrone offset by 0.753 mag.  That is the upper limit for unresolved close binaries
that are {\em bona fide} members of the cluster.
\label{cmd}
}

\begin{figure}
\plotone{ngc2482_cmd_new.eps} {Fig. \ref{cmd}.}
\end{figure}

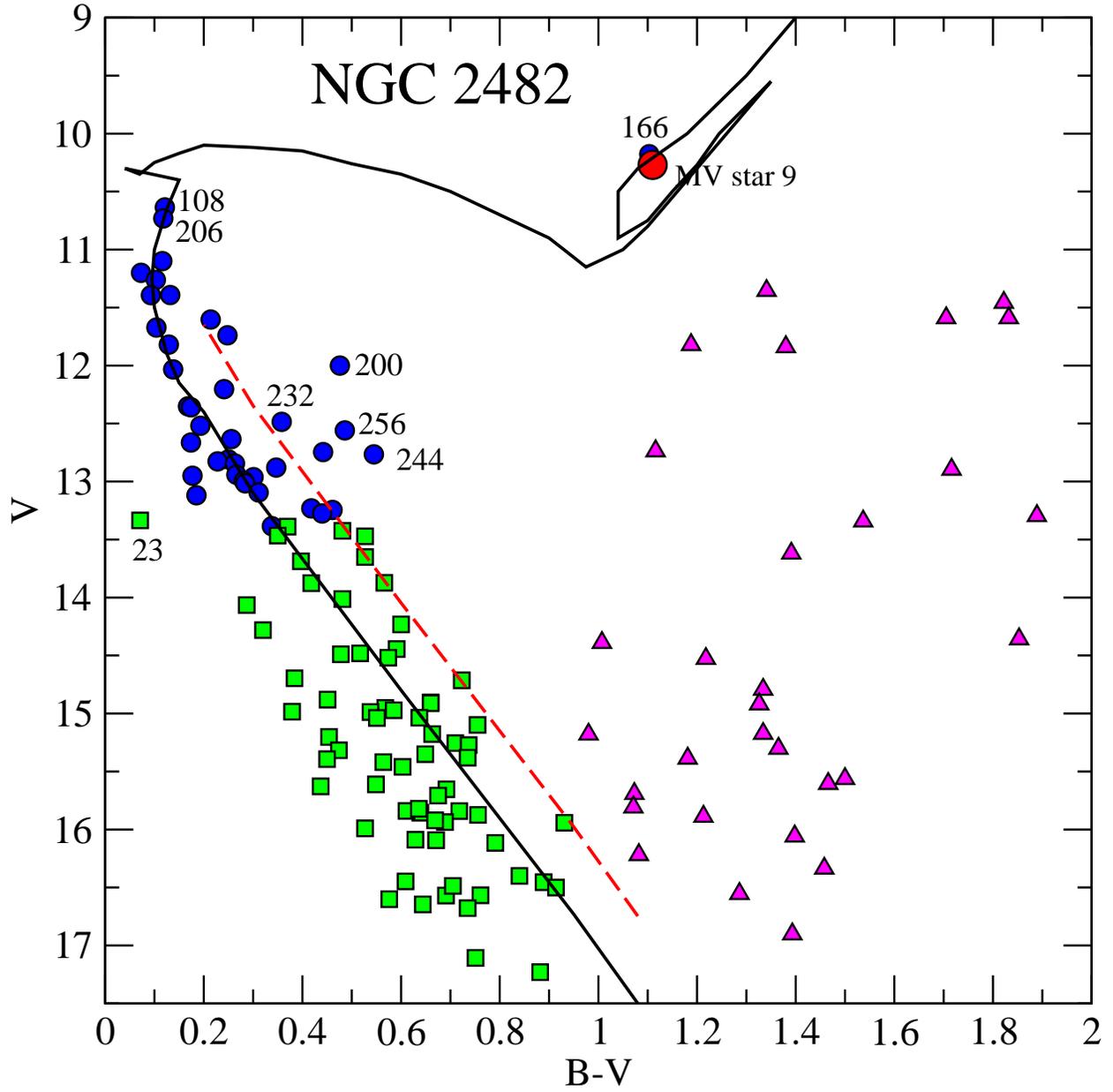
\end{document}